# Surface enhancement of superconductivity in tin

V.F. Kozhevnikov, M.J. Van Bael, W. Vinckx, K. Temst, C. Van Haesendonck, and J.O. Indekeu

Laboratorium voor Vaste-stoffysica en Magnetisme, Katholieke Universiteit Leuven, B-3001 Leuven, Belgium

The possibility of surface enhancement of superconductivity is examined experimentally. It is shown that single crystal tin samples with cold-worked surfaces represent a superconductor with a surface-enhanced order parameter (or negative surface extrapolation length $b$), whose magnitude can be controlled.

Properties of superconductors with surface enhancement of the order parameter, or with a negative value of the extrapolation length $b$ in a generalized boundary condition within Ginzburg-Landau theory [1], are a subject of long-standing discussions. Fink and Joiner [2] suggested, for the first time, that such a boundary condition should lead to an increase of the critical temperature in zero magnetic field. This means that the shape of the phase diagram for surface-enhanced superconductors may differ qualitatively from the classical one, for which both the bulk and the surface superconducting transitions have the same critical temperature, $T_c$, and there is no stable superconductivity above the critical point. We recall that the classical shape of the phase diagram is valid if the order parameter at the sample boundary has zero slope ($b = \infty$) or decreases ($b > 0$), which is appropriate for superconductor/vacuum or superconductor/normal-metal interfaces, respectively [1, 3]. Fink and Joiner have also shown experimentally that cold working the surface of a type-I superconductor ($In_{0.993}Bi_{0.007}$ alloy) increases the superconducting transition temperature, which they interpreted as a result of the surface enhancement; the observed shift in the critical temperature was about 0.02 K ($T_c$ for this alloy is 3.5 K).

Khlyustikov and Khaikin carried out an extensive experimental study of anomalous superconductivity at supercritical temperatures in pure metals, which is reviewed in Ref. 4. They found that mechanical treatment, such as bending and polishing, of carefully annealed (at a temperature only 0.1 K below the melting point) single-crystal samples of some metals (Sn, In, Nb, Re, and Tl) increases the critical temperature, whereas other metals (Al and Pb) do not show this effect. The maximum shift in the critical temperature, equal to 0.04 K, was observed in tin. However, in contrast to Fink and Joiner, Khaikin and Khlyustikov interpreted their observations as twinning-plane superconductivity occurring in the sample interior [5]. Twinning is the formation of two single-crystal regions (twins) so that the planar boundary between the twins is one of the crystallographic planes of the crystal. An important feature of twinning planes is that they do not involve stresses and therefore are not affected by annealing. Buzdin developed a Ginzburg-Landau theory of superconductivity in crystals with planar defects displaying enhancement of superconductivity, by incorporating negative *b*. The theoretical results are in agreement with the experimental data [4].

Indekeu and van Leeuwen [6] studied the consequences of surface enhancement in type-I superconductors within Ginzburg-Landau theory. An interface delocalization or "wetting" transition for surface-enhanced type-I superconductors was predicted. The transition is of first order for superconductors with low values of the Ginzburg-Landau parameter $\kappa$ (below 0.374), and of second order for the higher-$\kappa$ superconductors ($0.374 < \kappa < 1/\sqrt{2}$). In both cases nucleation of surface superconductivity in zero field occurs above the bulk critical temperature. It was also shown that the surface phase diagram for low-$\kappa$ superconductors has the same shape as for crystals with planar internal defects, regardless of the character of the defects (quantified by the transparency of the planar boundary to electrons) [7]. In other words, at low $\kappa$ the Ginzburg-Landau theory does not distinguish between a free surface, a grain boundary, or a twinning plane.

Thus we face a dilemma: the same experimental results can be interpreted in two principally different ways: either as an effect of *stresses* induced by surface treatment or as the effect of *stress-free* defects. For this reason it is interesting to examine which one of these interpretations is correct, or if both effects coexist, which one is dominant. Resolution of this dilemma constitutes the purpose of this paper. An additional motivation is associated with predictions [8, 9] that *surface* enhancement can yield a significant (up to a factor of ten) increase of the critical temperature for samples with dimensions of the order of or less than 1 μm. If this is confirmed, surface enhancement may find practical applications.

The experiments have been carried out with tin, because to our knowledge it exhibits the strongest anomalous superconductivity above $T_c$ [4]. The first samples we tested were cut from a high purity (99.9998%) tin foil of 0.1 mm thickness, fabricated by cold rolling (Alfar

Aesar). The sample size was 5 x 7 mm$^2$. After overnight annealing at 200 °C in vacuum (~10$^{-6}$ torr), DC magnetization was measured with a commercial SQUID magnetometer (Quantum Design, typical error for the temperature readings is ± 0.005 K) in a parallel magnetic field at temperatures above and below the bulk critical temperature $T_c$ ($T_c$ = 3.722 K [10]).

Results of the measurements for four super- and one subcritical isotherms are shown in Fig.1. The background magnetization from the sample holder was measured at 3.82 K and was subtracted from the data taken at lower temperatures. For samples 0.1 mm thick a distinct diamagnetic response was recorded starting from 3.78 K. The diamagnetic moment is several orders of magnitude (three orders for 3.74 K) stronger than that expected for fluctuation diamagnetism [3]. So we doubtlessly deal with the phenomenon of our interest, stable superconductivity above the bulk critical temperature, studied in Ref. 11 for single-crystal samples. However, in our case the effect is significantly stronger. In particular, the amplitude of the anomalous diamagnetic response reported in [11] for 3.76 K in terms of magnetization is about 5·10$^{-5}$ emu/cm$^3$; in our case it is 1.2·10$^{-3}$ emu/cm$^3$. Hysteresis for decreasing and increasing magnetic field is conspicuous for all temperatures, which indicates supercooling at nucleation of anomalous superconductivity, and is attributed to a first-order phase transition; this is in agreement with the observations of Khlyustikov and Khaikin [4, 11].

The slopes $\partial m/\partial H$ ($m$ is the magnetization in emu/cm$^3$, and $H$ is the magnetic field in Oe) for subcritical isotherms at low fields have the same value close to $1/4\pi$, as it should be for the Meissner state. The low-field slope for the supercritical isotherms decreases with temperature, allowing one to estimate the volume fraction of the anomalous superconducting phase. Such an estimate yields 0.5% at 3.78 K, 5% at 3.76 K, and up to 40% at 3.74 K.

Obviously, this amount of superconducting phase is too big to be consistent with superconductivity on the sample surface alone. As an additional check, the magnetization was measured after polishing of one and then also the other side of the sample. The magnetization exhibited an increase of about 10 to 20 % only, after polishing of a side. These results are definitely in favor of the bulk (sample interior) origin of the observed anomalies. On the other hand, so big a fraction of anomalous superconducting phase can hardly be consistent with twinning-plane superconductivity alone, in view of the delicate nature of twinning.

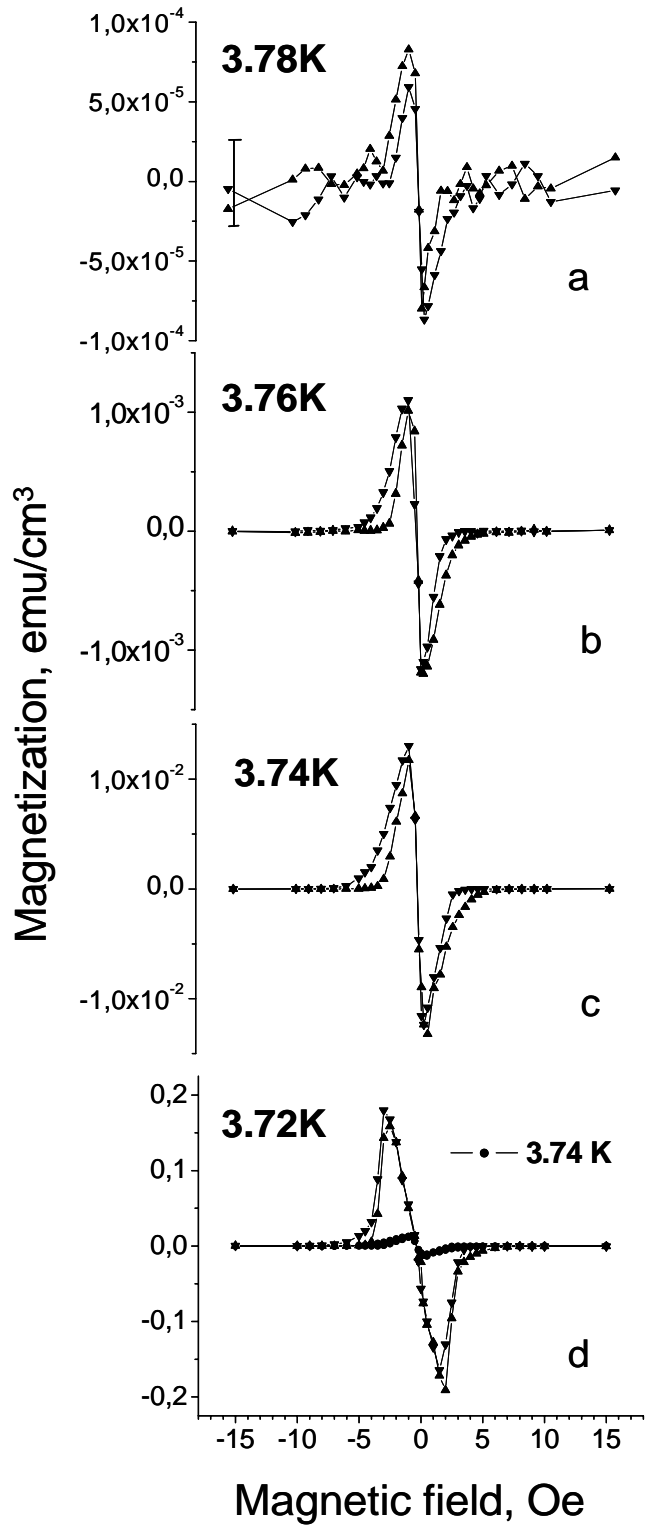

FIG. 1. Magnetization of the 0.1 mm thick tin foil in parallel magnetic field at super- (a, b, and c) and subcritical (d) temperatures. Down triangles are experimental points obtained for decreasing field, and up triangles are for increasing field. The error bar in (a) indicates average noise level, corresponding to $\pm 1\cdot 10^{-7}$ emu. Points for temperature 3.74 K (solid circles) are repeated in (d) for comparison.

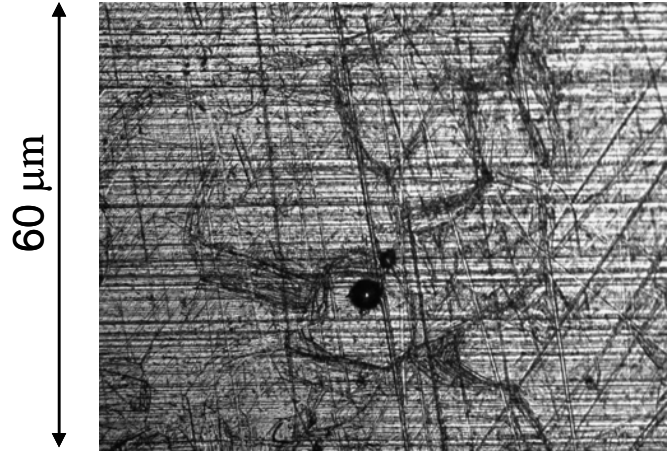

FIG. 2. A photograph of the surface of the tin foil. The scratches are traces of cold rolling.

Examining the foil samples with an optical microscope (Jenavert, Sypac s.c.s.), we noticed that the cold-rolled foil consists of flake-like grains with a typical size of the order of 10 µm, and a thickness of the order of 1 µm. A photograph of the foil surface is shown in Fig.2. The grains appear to be rather weakly bonded, since the foil can be disintegrated by agitation in an ultrasonic bath. Therefore, the real surface area is the sum of the grain surfaces, which is much larger than the area of the sample outer surface. Measurements of the magnetization of an ultrasonically disintegrated foil (one-hour exposure in a commercial ultrasonic bath Branson 5200; the sample was in a beaker with acetone at room temperature) yielded an anomalous signal similar to that shown in Fig.1, but with a magnitude smaller by about a factor of three. Foils annealed for a longer time (70 hours) and at a higher temperature (230 °C) also yielded an anomalous magnetization lower by about a similar factor. These observations contradict the twinning-plane interpretation and favor the stressed defect (interior and exterior) origin of the effect. However, the residual diamagnetism recorded in the annealed or in the disintegrated foil samples *can* be due to twinning planes or other low-stress defects. Therefore subsequent experiments were performed on samples with minimal amounts of such defects, that is, single-crystal tin samples.

The single-crystal samples (Alfa Easar, cast from 99.9999 pure tin) had a disk shape about 7 mm in diameter and 1 mm in thickness. We worked with two samples. Sample #1 was first annealed at 230 °C for 100 hours and then polished on both sides with a silicon carbide grinding paper # 4000 (Struers S.A.S., grain size 5 μm). Sample # 2 was first polished, then annealed and then polished again. DC magnetization has been measured after each of these steps. The chosen annealing temperature, 230 °C, turned out to be optimal for providing sufficiently close proximity to the melting point (232°C) and minimal risk of sample overheating at the temperature stabilization.

The anomalous magnetization of sample #1 in the annealed state was not noticeable down to 3.76 K. At temperatures 3.76 K and 3.74 K, its amplitude was $2 \cdot 10^{-5}$ and $2 \cdot 10^{-4}$ emu/cm$^3$, respectively, which is two orders of magnitude lower than that observed for the foils. After polishing, the magnetization of this sample increased by a factor of 5 to 6.

The magnetization data obtained for sample # 2 are shown in Fig. 3. After annealing the magnitude of the anomaly dropped by a factor of ten; then it was completely restored by repolishing.

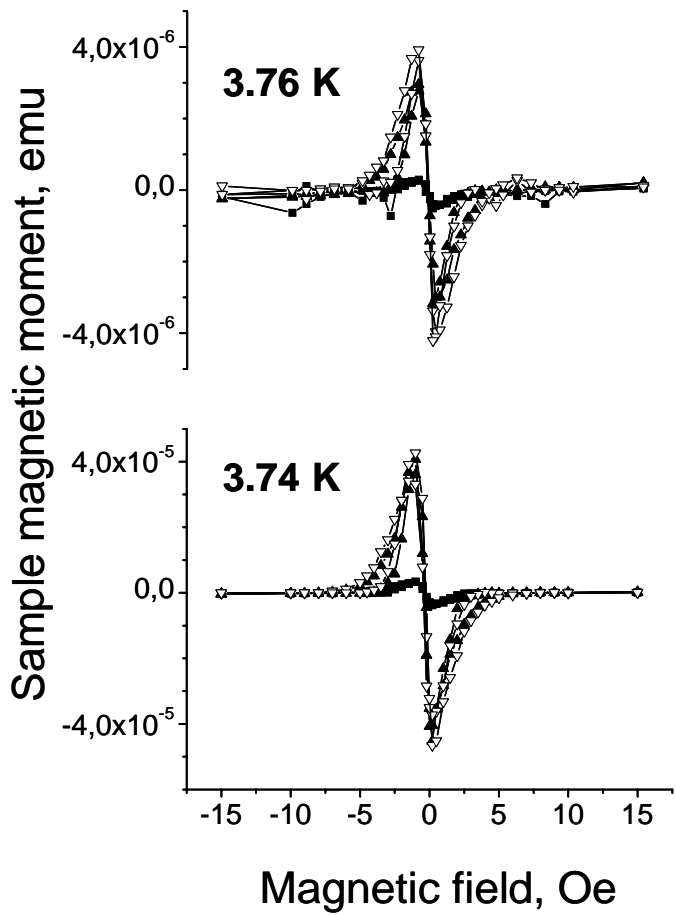

FIG. 3. Magnetic moment of the polished (closed triangles), annealed (closed squares), and repolished (open triangles) single-crystal sample

We also checked if annealing-polishing procedures affect the bulk transition temperature, via measurements of the temperature dependence of the magnetization near $T_c$ in a field of 0.5 Oe with polished sample #1 and annealed sample #2. Results are shown in Fig.4. The magnetization at the foot of the transition is shown in the insert on enlarged scale. There is no visible change in the bulk transition temperature, whereas the foot does change both in magnitude and in temperature width: in the polished sample the foot is higher and wider. This is an additional confirmation of surface enhancement of superconductivity by cold working the sample surface.

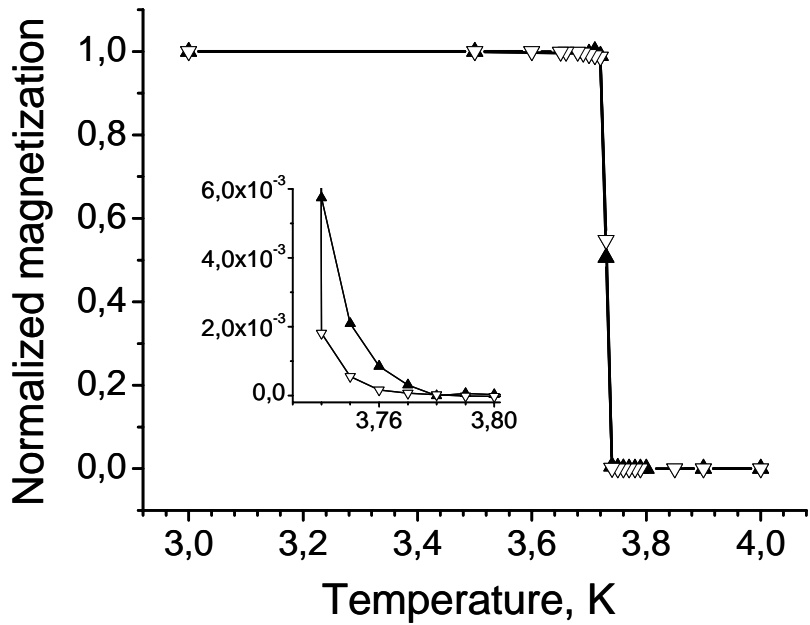

FIG. 4. Normalized magnetization (with reversed sign) for the annealed (open triangles) and polished (solid triangles) single crystal tin samples at 0.5 Oe. The inset shows the magnetization over the foot of the transition.

Thus we arrive at the following conclusions. (a) The anomalous superconductivity in both poly- and single-crystal tin samples is mainly hosted by outer and internal *stressed* defects. (b) Surface superconductivity in tin can be induced using mechanical polishing (surface cold working). (c) The nucleation of surface superconductivity occurs as a first-order phase transition. (d) Single-crystal tin samples with polished surfaces represent a superconductor with surface-enhanced order parameter and, correspondingly, with negative extrapolation length *b*. (e) The enhancement strength (or magnitude of *b*) can be *controlled,* for instance, by manipulating the abrasive grain size and the annealing parameters (time and temperature). This point requires further study, however.

We believe that the residual anomalous magnetization measured in the single-crystal samples could be further reduced by annealing at closer proximity to the melting point. However, it is possible that this *minor* part of the anomaly comes from low-stress defects, such as twinning planes. Finally, the fabricated surface-enhanced tin samples provide a reproducible basis for setting up an experimental verification of the theoretically predicted interface delocalization phenomena in type-I superconductors.

**Acknowledgements:** This research has been supported in part by the Research Council of the K.U.Leuven (fellowship F/03/066 for V.F.K.) and by Project G.0237.05 of the Fund for Scientific Research of Flanders.